\documentclass{article}
\usepackage{ijcai19}

\usepackage{times}
\usepackage{soul}
\usepackage{url}
\usepackage[hidelinks]{hyperref}
\usepackage[utf8]{inputenc}
\usepackage[small]{caption}
\usepackage{graphicx}
\usepackage{amsmath}
\usepackage{booktabs}
\usepackage{color}
\usepackage{epsfig}
\usepackage[normalem]{ulem}
\usepackage{enumerate}
\usepackage{multirow}

\def\cL{{\cal L}}

\newcommand{\fig}[1]{Figure~\ref{#1}}

\newcommand{\Ln}{{\Lambda_\nu}}

\newcommand{\ajin}{{a_{j\to i}^\nu}}
\newcommand{\aijn}{{a_{i\to j}^\nu}}

\newcommand{\aljn}{{a_{l\to j}^\nu}}

\newcommand{\bjin}{{b_{j\to i}^\nu}}
\newcommand{\bijn}{{b_{i\to j}^\nu}}

\newcommand{\bljn}{{b_{l\to j}^\nu}}
\newcommand{\brjn}{{b_{r\to j}^\nu}}

\newcommand{\Ljn}{{\Lambda_j^\nu}}

\newcommand{\ejins}{{\eta_{ji}^{\nu*}}}

\newcommand{\cost}{T}
\newcommand{\tim}{{\tilde t}}

\title{Learning the optimally coordinated routes from the statistical mechanics of polymers}

\author{
Hao Liao $^1$
\and
Xingtong Wu$^1$\and
Mingyang Zhou$^1$\And
Chi Ho Yeung$^2$
\affiliations
$^1$National Engineering Laboratory for big data computing systems, Guangdong Province Key Laboratory of Popular High Performance Computers, College of Computer Science and Software Engineering, Shenzhen University\\
$^2$Department of Science and Environmental Studies, The Education University of Hong Kong, Hong Kong, PR China
\emails
haoliao@szu.edu.cn
}

\begin{document}

\maketitle

\begin{abstract}
Many major cities suffer from severe traffic congestion. Road expansion in the cites is usually infeasible, and an alternative way to alleviate traffic congestion is to coordinate the route of vehicles. Various path selection and planning algorithms are thus proposed, but most existing methods only plan paths separately and provide un-coordinated solutions. Recently, an analogy between the coordination of vehicular routes and the interaction of polymers is drawn; the spin glass theory in statistical physics is employed to optimally coordinate transportation routes. To further examine the advantages brought by path coordination, we incorporate the link congestion function developed by the Bureau of Public Roads (BPR) into the polymer routing algorithm. We then estimate in simulations the traveling time of all users saved by the polymer-BPR algorithm in randomly generated networks and real transportation networks in major cities including London, New York and Beijing. We found that a large amount of traveling time is saved in all studied networks, suggesting that the approach inspired by polymer physics is effective in minimizing the traveling time via path coordination, which is a promising tool for alleviating traffic congestions.
\end{abstract}

\section{Introduction}

Traffic congestion is a recurrent severe problem in many major cities in the world ~\cite{song2016deeptransport}. In many cases, popular routes are highly congested, leading to a huge time cost for road users while leaving some longer routes potentially under-used. One effective way to mitigate traffic congestion is to coordinate the paths of all vehicles ~\cite{chardaire2005solving}, i.e. to assign the paths of each individual vehicle in the network simultaneously in a coordinated manner, in order to achieve a global objective ~\cite{wu2006transport}. Nevertheless, it is not an easy task, and most existing routing algorithms only optimize static routes or steady flows separately ~\cite{yang2016approximation}. For instance, navigation applications usually suggest the shortest path to individual vehicles based on the current traffic condition~\cite{xu2018effective}, which is an uncoordinated routing strategy by which many vehicles may to through the same route simultaneously, leading to traffic congestion.

In recent years, statistical physics has been applied to optimize routes in transportation networks. Examples include the identification of routes in Steiner trees and the optimization of paths from multiple origins to a universal destination~\cite{yeung2012competition}. Recently, transportation routes have been mapped to interacting polymers~\cite{galina1988some,yeung2013physics,saad2014physics}, of which the two polymer ends correspond to the origin and the destination of the route of individual road user. Repulsion is introduced in order to minimize the overlap between the polymers on the network. A message-passing algorithm is then derived based on the techniques developed in the study of spin glasses~\cite{mezard1987spin,sanghavi2009message}, which optimally arranges the interacting polymers on the network, leading to an algorithm which optimally coordinate the routes to achieve a global objective.

In this paper, instead of introducing a repulsive interaction between the paths of vehicles, we incorporate the link congestion function developed by the Bureau of Public Roads (BPR)~\cite{bpr1964traffic,stefanello2016using,grunitzki2016combining,zhang2017optimal} into the polymer routing algorithm~\cite{saad2014physics}. Since the BPR link congestion function corresponds to the traveling time along a road given its amount of traffic, optimizing the BPR function on the whole network is equivalent to minimizing the total traveling time for all users using the network. Hence, our algorithm which incorporates the BPR function in the polymer routing algorithm is thus an algorithm which minimizes the total traveling time for all road users.

Compared to existing local greedy algorithms, our global optimization algorithm which we call polymer-BPR can allocate multiple interacting paths at the same time and minimize a global cost. Algorithmically, our polymer-BPR is a message-passing algorithm ~\cite{swoboda2017message}, and its distributive nature is beneficial for implementation in navigation applications, compared to centralized optimization algorithms such as dynamical linear programming ~\cite{xie2004self,zhu2006dynamic}. By applying our algorithm on random regular graphs~\cite{liu2011large}, the London Underground network, the New York Road network, and the Beijing Rail Transit network, we found that the traveling time estimated by the BPR function and obtained by our polymer-BPR algorithm is much shorter than that obtained when all users commute by their shortest path.


\begin{figure*}[htb]
\centering\includegraphics[height=1.77in]{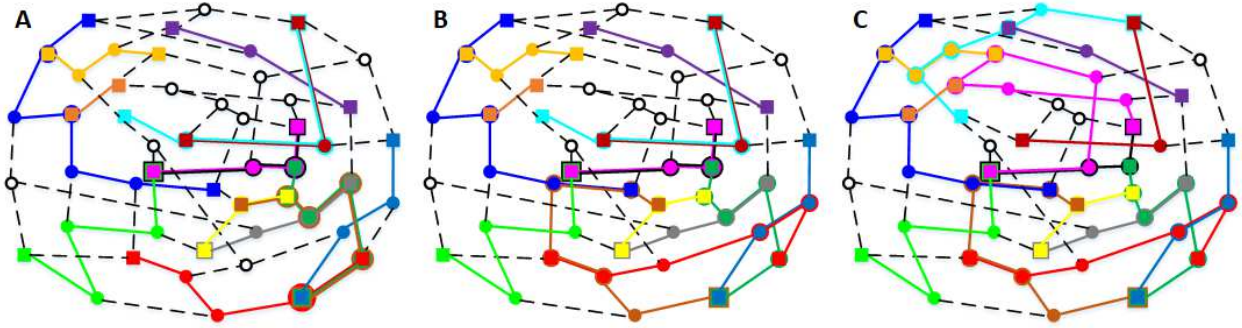}
\caption{
The path configurations identified by (a) the Dijkstra algorithm, (b) the polymer routing algorithm, and (c) the polymer-BPR algorithm on a regular random network with $N=50$ nodes, node degree $k=3$ and $M=15$ source-destination pairs. The value of the total traveling time $\cost=164.35$, $99.15$ and $86.9$ respectively. The path of each vehicle is represented by an individual color; when multiple vehicle pass via the same nodes and edges, their colors overlap. The node size is proportional to the amount of traffic passing through it. Square nodes represent the origin or the destination of each vehicle. Compared with the configuration obtained by the Dijkstra algorithm, the polymer-BPR algorithm reduces $\cost$ by almost $50\%$. The polymer-BPR algorithm also lead to a lower $\cost$ when compared to the original polymer algorithm. These results indicate that the polymer-BPR algorithm can be used to coordinate vehicle path for saving the global traveling time. The number of nodes and edges shared by multiple vehicles in the path configuration obtained by the three algorithms are shown in Table~\ref{tab:share}.
 }
\label{fig:ex}
\end{figure*}

\newcommand{\tabincell}[2]{\begin{tabular}{@{}#1@{}}#2\end{tabular}}
\begin{table*}
\centering
\begin{tabular}{p{85pt}p{40pt}p{40pt}p{60pt}p{5pt}p{40pt}p{40pt}p{60pt}}
\toprule
\multirow{2}{3cm}{Number of sharing vehicles} & \multicolumn{3}{c}{Number of {\bf edges}} && \multicolumn{3}{c}{Number of {\bf nodes}} \\
\cmidrule{2-4}\cmidrule{6-8}
 & \multicolumn{1}{c}{Dijkstra} & \multicolumn{1}{c}{Polymer} & \multicolumn{1}{c}{Polymer-BPR} && \multicolumn{1}{c}{Dijkstra} & \multicolumn{1}{c}{Polymer} & \multicolumn{1}{c}{Polymer-BPR} \\
\midrule
\multicolumn{1}{c}{0}&\multicolumn{1}{c}{\textbf{38}}&\multicolumn{1}{c}{33}&\multicolumn{1}{c}{24}&&\multicolumn{1}{c}{\textbf{10}}&\multicolumn{1}{c}{9}&\multicolumn{1}{c}{5} \\
\hline
\multicolumn{1}{c}{1}&\multicolumn{1}{c}{\textbf{27}}&\multicolumn{1}{c}{33}&\multicolumn{1}{c}{46}&&\multicolumn{1}{c}{24}&\multicolumn{1}{c}{\textbf{19}}&\multicolumn{1}{c}{21} \\
\hline
\multicolumn{1}{c}{2}&\multicolumn{1}{c}{8}&\multicolumn{1}{c}{9}&\multicolumn{1}{c}{\textbf{5}}&&\multicolumn{1}{c}{\textbf{9}}&\multicolumn{1}{c}{19}&\multicolumn{1}{c}{22} \\
\hline
\multicolumn{1}{c}{3}&\multicolumn{1}{c}{2}&\multicolumn{1}{c}{0}&\multicolumn{1}{c}{\textbf{0}}&&\multicolumn{1}{c}{6}&\multicolumn{1}{c}{3}&\multicolumn{1}{c}{\textbf{2}} \\
\hline
\multicolumn{1}{c}{4}&\multicolumn{1}{c}{0}&\multicolumn{1}{c}{0}&\multicolumn{1}{c}{0}&&\multicolumn{1}{c}{1}&\multicolumn{1}{c}{0}&\multicolumn{1}{c}{\textbf{0}} \\
\bottomrule
\end{tabular}
\caption{The number of nodes and edges shared by multiple vehicles in the path configurations identified by the Dijsktra algorithm, the polymer routing algorithm and the polymer-BPR algorithm.}
\label{tab:share}
\end{table*}

\section{Models and methods}
To examine how transportation routes are coordinated by various algorithms, we consider $M$ vehicles commuting on a network with $N$ nodes. Each node $i$ on the network is connected to $k_i$ neighboring sites. The origin and the destination of each vehicle $\nu$ are nodes on the network. We denote the variable $\sigma_{(ij)}^\nu=1$ when vehicle $\nu$ commutes using the edge $(ij)$ connecting node $i$ and $j$, and $\sigma_{(ij)}^\nu=0$ otherwise. The traffic flow $I_{(ij)}$ on the edge $(ij)$ is thus the number of vehicles using the edge, i.e.
\begin{align}
I_{(ij)}=\sum_\nu \sigma_{(ij)}^\nu
\end{align}
To estimate the total traveling time of all vehicles, we will adopt the link congestion function developed by the Bureau of Public Roads (BPR)~\cite{bpr1964traffic}, the then U. S. Federal Highway Administration. We will call this function the \emph{BPR function} throughout the paper, which gives the traveling time of a vehicle through an edge $(ij)$ with traffic flow $I_{(ij)}$ to be
\begin{align}
t_{(ij)}(I_{(ij)}) = \tim_{(ij)}\left[1+0.15\left(\frac{I_{(ij)}}{C_{(ij)}}\right)^4\right],
\end{align}
where $\tim_{(ij)}$ and $C_{(ij)}$ are the free-flow travel time and the capacity on the edge $(ij)$. In this case, one can define a cost function $\phi(I_{(ij)})$ on an edge $(ij)$ with traffic flow $I_{(ij)}$, which correspond to the total passage time of all vehicles using the edge
\begin{align}
\label{eq:BPR}
\phi(I_{(ij)}) = I_{(ij)} t(I_{(ij)}) = I_{(ij)} \tim_{(ij)}\left[1+0.15\left(\frac{I_{(ij)}}{C_{(ij)}}\right)^4\right].
\end{align}
The total traveling time $\cost$ for all vehicles in the network who travel from their respective origin to destination is given by
\begin{align}
\cost = \sum_{(ij)} \phi(I_{(ij)}),
\end{align}

With the BPR function, we can compute the estimated traveling time when vehicles go by their respective shortest paths without coordination, or their paths are coordinated by congestion-aware path planning algorithms. Here, we will describe the routing algorithms that are studied in the present work.

\subsection{To identify the shortest path – The Dijkstra algorithm}

The shortest path algorithm proposed by Dijkstra in 1959~\cite{dijkstra1959note} is the benchmark algorithm for dealing with routing problems. The algorithm uses a breadth-first search strategy to obtain the shortest path between a specific node and all other nodes in the network in a single run. The Dijkstra algorithm is simple to implement and its computational complexity is $O(N \log N)$. Nevertheless, since only the shortest paths are obtained by the Dijkstra algorithm, transportation routes obtained by the algorithm are not coordinated and may lead to severe traffic congestion.

\subsection{The polymer routing algorithm}

The \emph{polymer routing algorithm} was derived in~\cite{yeung2013networkinga}, which utilizes the analogy between repulsive interacting polymers and overlap-suppressing transportation routes, to achieve congestion-aware coordinated path configurations in transportation networks. By applying techniques developed in the studies of spin glasses and disorder systems, a distributive message-passing algorithm capable to identify the low-energy states of a system of interacting polymers is derived in~\cite{yeung2013physics}, which is equivalent to a configuration of coordinated transportation paths on the network.

Given a cost function $\phi(I_{(ij)})$ defined on an edge $(ij)$ with current $I_{(ij)}$, the polymer routing algorithm which involves only edge cost works by passing messages from each node $i$ to its neighbor $j$ for each index $\nu$ until convergence. The two messages $\ajin$ and $\bjin$ from node $i$ to node $j$ for index $\nu$ are given by
\begin{align}
\label{eq:a2}
\ajin =
\begin{cases}
\displaystyle
\min_{l\in \cL_j\backslash \{i\}}\left[\aljn\right]+ \phi'(\ejins)
\\
\quad\quad-\min\left[0, \min_{\substack{l,r\in \cL_j\backslash\{i\}\\l\neq r}}\left[\aljn+\brjn\right]\right],
\\
&\hspace{-1.5cm} \Ljn=0
\\
\displaystyle
-\min_{l\in \cL_j\backslash \{i\}}\left[\bljn\right]+ \phi'(\ejins),
&\hspace{-1.5cm} \Ljn=1
\\
\displaystyle
\infty,
&\hspace{-1.5cm} \Ljn=-1
\end{cases}
\\
\label{eq:b2}
\bjin =
\begin{cases}
\displaystyle
\min_{l\in \cL_j\backslash \{i\}}\left[\bljn\right]+ \phi'(\ejins)
\\
\quad\quad-\min\left[0, \min_{\substack{l,r\in \cL_j\backslash\{i\}\\l\neq r}}\left[\aljn+\brjn\right]\right],
\\
&\hspace{-1.5cm} \Ljn=0
\\
\displaystyle
\infty,
&\hspace{-1.5cm} \Ljn =1
\\
\displaystyle
\phi_E'(\ejins),
&\hspace{-1.5cm} \Ljn=-1
\end{cases}
\end{align}
The value of $\ejins$ is given by
\begin{align}
\label{eq:ejins}
\ejins = \frac{1}{M} + \sum_{\mu\neq\nu}\sigma_{ji}^\mu,
\end{align}
such that $\sigma_{ji}^\nu$ is given by
\begin{align}
\label{eq:sigmaji}
\sigma_{ji}^\nu &=
\delta_{\Ln, 1}\Theta\left(\min_{l\in \cL_j\backslash \{i\}}\left[\bljn\right]-\bijn\right)
\nonumber\\
&+ \delta_{\Ln, -1}\Theta\left(\min_{l\in \cL_j\backslash \{i\}}\left[\aljn\right]-\aijn\right)
\nonumber\\
&+ \delta_{\Ln, 0}\Theta\left(\min\left[0, \min_{\substack{l,r\in \cL_j\backslash\{i\}\\l\neq r}}\left[\aljn+\brjn\right]\right]
\right.
\nonumber\\
&
-\min\left[\aijn\!+\!\min_{l\in \cL_j\backslash \{i\}}\left[\bljn\right], \bijn\!+\!\min_{l\in \cL_j\backslash \{i\}}\left[\aljn\right]
\right]\Bigg),
\end{align}
which corresponds to the optimized flow of the edge $(ij)$ for vehicle $\nu$ after the convergence of messages: $\sigma_{ji}^\nu=1$ if vehicle $\nu$ passes the edge and $\sigma_{ji}^\nu=0$ otherwise.

\subsection{The polymer-BPR algorithm}

In the original work of the polymer routing algorithm, the cost function on the edge is given by $\phi(I)\propto I^\alpha$. The cases with $\alpha=2$ are used to represent repulsive polymers, as $\phi(I)$ increases more than linearly as $I$ increases and the case with a large number of polymers on the same edge is penalized. The results in~\cite{yeung2013physics} show that routes are well distributed in this case, but the values of the total cost function do not correspond to a physical quantity of interest. In the present work, we will integrate the cost function in Equation \ref{eq:BPR} to the polymer routing algorithm, which we call the \emph{polymer-BPR} algorithm. This algorithm aims to minimize the total traveling time of all vehicles. The results are compared to path configurations obtained by other benchmarking algorithms, which lead to insights into the amount of traveling time which can be saved by coordinating vehicle paths.


\section{Data description and evaluation metrics}

In the subsequent sections, we will use the generated random regular graphs, the London Underground network\footnote{http://www.tfl.gov.uk}, the Beijing Rail Transit network\footnote{http://www.smartcity-competition.com.cn/} and the New York road network\footnote{https://www.nature.com/articles/sdata201646} to verify the performance of our algorithm. The descriptions of these real graphs are as follows:

\begin{figure*}
\centering\includegraphics[height=3in]{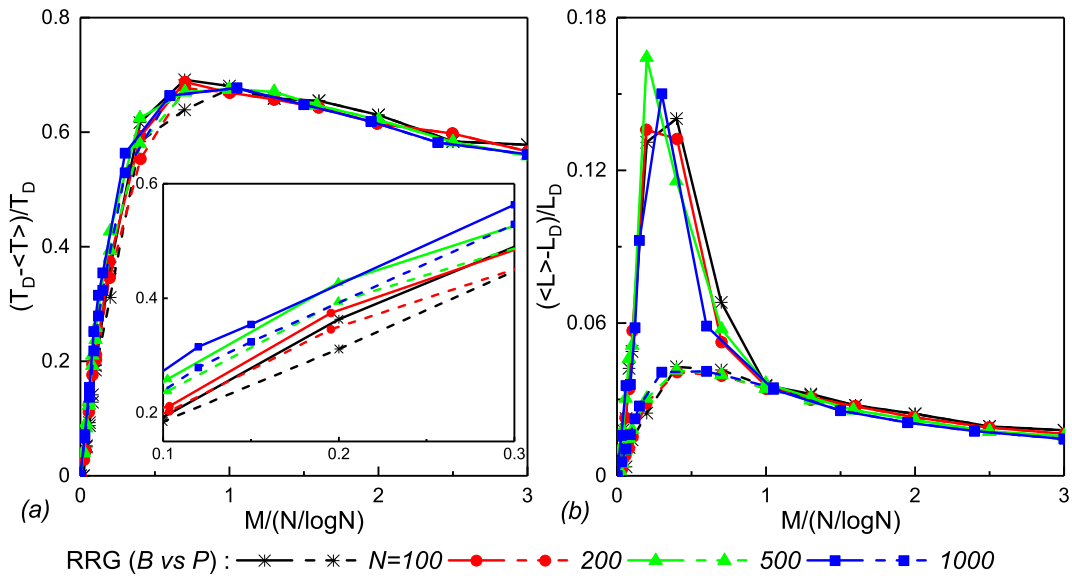}
\caption{
(a) The fractional difference $(\cost_{D}-\langle \cost\rangle)/\cost_{D}$ in the traveling time, and (b) the fractional difference $(\langle L\rangle-L_{D})/L_{D}$ in the path length, between the path configurations identified by the polymer-BPR algorithm (as well as the polymer routing algorithm with $\alpha=2$) and the Dijsktra algorithm, as a function of $M/(N/\log N)$ which corresponds to the re-scaled number of vehicles. The subscripts B, P and D stand for the polymer-BPR algorithm, the polymer routing algorithm and the Dijkstra algorithm respectively. The value of $L_D$ in (b) corresponds to the average shortest path in the graphs. All simulation results are averaged over 2000 realizations of random regular graphs with network size $N=100, 200, 500, 1000$ and node degree $k=3$.
 }
\label{fig:sim}
\end{figure*}

\begin{figure}
  \centering
  \includegraphics[width=3.3in]{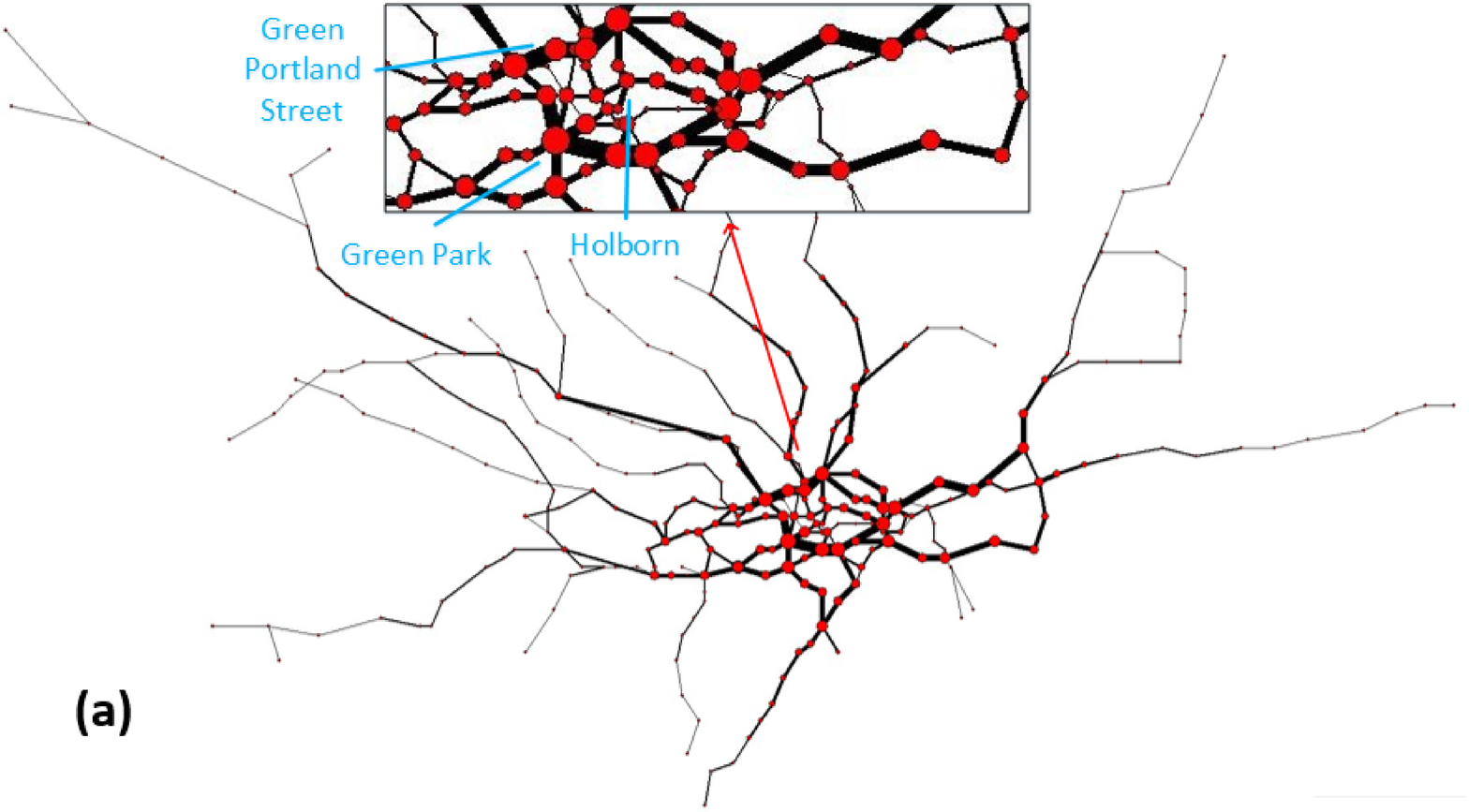}
  \includegraphics[width=3.3in]{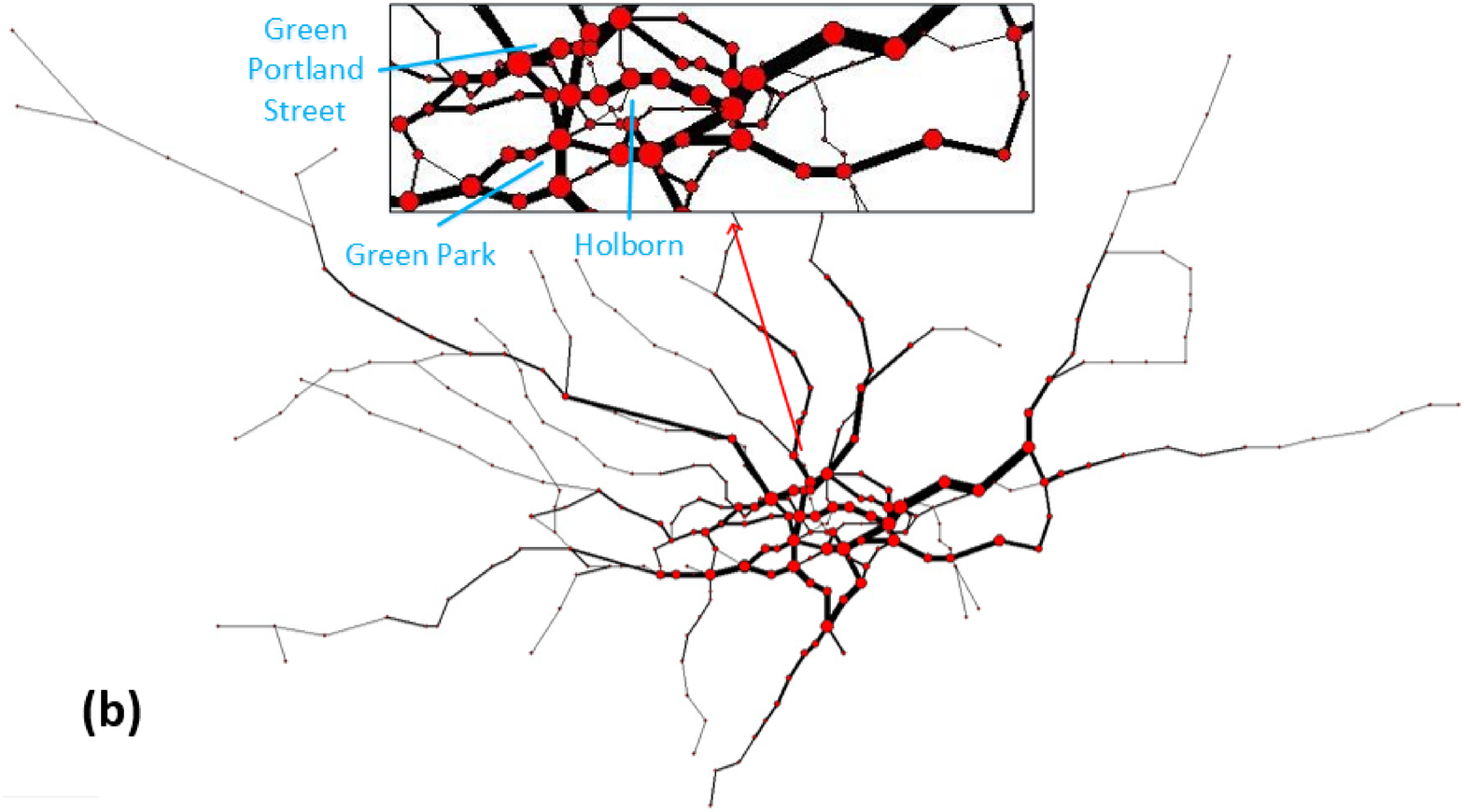}
  \caption{Traffic on the London subway network optimized by the (a) polymer routing algorithm, and (b) the polymer-BPR algorithm. A total of 218 real passenger source–destination pairs are optimized, corresponding to $5\%$ of the data recorded by the Oyster card system between 8:30 AM and 8:31 AM on one Wednesday in November 2009 (35). Red nodes correspond to stations with nonzero traffic. The size of each node and the thickness of each edge are proportional to traffic through them. Insets: Enlarged views of the central region. Nodes are drawn according to their geographic position.}
  \label{fig:ld}
\end{figure}

\begin{enumerate}
\item In our generated random regular graphs with $N$ nodes, each node has the same degree $k$. The graph size used in the experiment includes $N =100$, $200$, $500$ and $1000$, with $k=3$.
\item The London Underground network consists of 275 stations, each of which represents a node in the graph, and the number of edges, i.e. direct connection between stations, is 373. We also used the real origin-destination pairs of passengers recorded by the Oyster card system, between 8:24 am - 8:40 am on one of the mornings in November 2009.
\item The New York Road network contains 3642 nodes (i.e. junctions) and 5904 edges (i.e. roads). A part of the network with far-reaching geographical coordinates is removed from the network used in the simulations, reducing the difference in edge weights. In the experiment, we have set $\tim_{(ij)}$ and $C_{(ij)}$ in the BPR function Equation~(\ref{eq:BPR}) to be the length and the number of lanes on the road $(ij)$ in the real network to have an accurate representation of the efficiency of individual road.
\item The Beijing Rail Transit network is the first subway system in China. There is an average daily passenger flow of 10 million passengers. It is one of the busiest subway systems in China. In the experiment, the network we investigated is composed of 19 subway lines, and the number of stations is 288 with 328 connections between stations. The network data was obtained from Beijing Subway - Baidu Map\footnote{http://map.baidu.com/?subwayShareId=beijing,131}.
\end{enumerate}

Other than the New York road network, we have set $\tim_{(ij)}=1$ and $C_{(ij)}=1$ in the BPR function Equation~(\ref{eq:BPR}) for the other three types of network.

\section{Results}

\begin{table*}
\centering
\begin{tabular}{p{75pt}p{90pt}p{90pt}p{5pt}p{85pt}p{85pt}}
\toprule
&\multicolumn{2}{c}{Polymer-BPR versus Dijkstra} & & \multicolumn{2}{c}{Polymer-BPR versus Polymer ($\gamma=2$)} \\
\cmidrule{2-3}\cmidrule{5-6}
& \multicolumn{1}{c}{$\frac{ E_B-E_D }{E_D}$} & \multicolumn{1}{c}{$\frac{ L_B-L_D }{L_D}$} && \multicolumn{1}{c}{$\frac{ E_B-E_P }{E_P}$} & \multicolumn{1}{c}{$\frac{ L_B-L_P }{L_P}$} \\
\midrule
\tabincell{c}{London Underground\\Network} & \tabincell{c}{$-71.7\%$\\$[-93.6\%,-38.1\%]$} & \tabincell{c}{$+19.6\%$\\$[13.1\%,28.7\%]$} && \tabincell{c}{$-17.6\%$\\$[-31.4\%,-4.0\%]$} & \tabincell{c}{$+12.2\%$\\$[6.8\%,17.5\%]$} \\
\tabincell{c}{NewYork Road\\Network} & \tabincell{c}{$-82.3\%$\\$[-98.1\%,-61.1\%]$} & \tabincell{c}{$+21.8\%$\\$[17.0\%,29.1\%]$} && \tabincell{c}{$+1.6\%$\\$[-5.3\%,5.6\%]$} & \tabincell{c}{$+11.6\%$\\$[8.1\%,16.2\%]$} \\
\tabincell{c}{Beijing Rail Transit\\Network} & \tabincell{c}{$-68.4\%$\\$[-91.0\%,-37.3\%]$} & \tabincell{c}{$+10.8\%$\\$[5.3\%,16.0\%]$} && \tabincell{c}{$-16.8\%$\\$[-40.7\%,-5.7\%]$} & \tabincell{c}{$+7.1\%$\\$[3.1\%,10.8\%]$} \\
\bottomrule
\end{tabular}
\caption{
The fractional differences in path length and cost $\cost$ between the path configurations identified by the polymer-BPR algorithm and the Dijsktra algorithm (or the polymer routing algorithm with $\gamma=2$), on the London Underground network, the New York road network and the Beijing Rail Transit network respectively. The origin-destination pairs on the London Underground network are taken from 1\% of the data recorded by Oyster Card in one of the mornings in November 2009. On the other hand, 10 groups of 40 random origin-destination pairs are generated on the New York road networks, and 50 groups of 30 random origin-destination pairs are generated on the Beijing Rail Transit networks. The values $a$ and $b$ in $[a, b]$ correspond to the upper or upper obtained value among the trials.}
\label{tab:energyDiff}
\end{table*}

We first show the path configuration identified by the Dijsktra algorithm, the polymer routing algorithm and the polymer-BPR algorithm for 15 identical origin-destination pairs on the same random regular graph with $N=50$ and $k=3$ in Figure~\ref{fig:ex}. The corresponding number of nodes and edges which are shared by the multiple vehicles are shown in Table~\ref{tab:share}. As we can see in Figure~\ref{fig:ex}(a) and Table~\ref{tab:share}, the path configuration identified by the Dijsktra algorithm has the largest number of empty nodes and edges, but it also has the largest number of triply-occupied nodes and edges, as well as a node occupied by 4 vehicles, which may lead to severe congestion. In reality, the self-decisive and independent planning of the shortest path may result in highly overlapping paths, leading to a high load on specific nodes and edges and hence a large total traveling time $\cost$, which may be a common origin of congestion in real transportation networks.

We show in Figure~\ref{fig:ex}(b) the path configuration identified by the polymer routing algorithm. As we can see in Table~\ref{tab:share}, the number of empty nodes and edges is lower than those found by the Dijsktra algorithm, but the number of nodes and edges occupied by multiple vehicles is also lower, which leads to a shorter global traveling time $\cost$ as estimated by the BPR function. This implies that the global traveling time $\cost$ can be reduced by coordinating vehicle paths. We then show in Figure~\ref{fig:ex}(c) the path configuration identified by our purposed polymer-BPR algorithm. As we can see in Table~\ref{tab:share}, the number of doubly and triply occupied edges and triply occupied nodes are the least among the three algorithms, leading to the shortest global traveling time $\cost$, even shorter than that found by the polymer routing algorithm. These results suggest that the polymer-BPR algorithm can effectively coordinate the paths of vehicle to minimize their total traveling time.

To quantitively show the benefit in saving traveling time by the polymer-BPR algorithm, we show in Figure~\ref{fig:sim}(a) the fractional difference $(\cost_D-\cost_B)/\cost_D$ between the global traveling time in the path configurations identified by the Dijsktra and the polymer-BPR algorithm, as a function the re-scaled number of vehicles $M/(N/\log N)$. As we can see, $(\cost_D-\cost_B)/\cost_D > 0.5$ for a large range of $M/(N/\log N)$, which suggests that the polymer-BPR algorithm is capable in coordinating path and saving traveling time by more than $50\%$ when compared to the case when users travel via their shortest path. For comparison, we also show the fractional difference between the global traveling time found by the Dijsktra and the polymer routing algorithm in Figure~\ref{fig:sim}(a). As we can see in the inset, $(\cost_D-\cost_B)/\cost_D > (\cost_D-\cost_P)/\cost_D $, suggesting that the polymer-BPR algorithm performs better than the polymer algorithm in saving traveling time as estimated by the BPR function.

Indeed, the global traveling time is saved by reducing wasted time in congestion, when vehicles travel via longer paths. In Figure~\ref{fig:sim}(b), we show the fractional difference between the path length in the configurations identified by the polymer-BPR and the polymer algorithm, with that of the shortest path configurations. As we can see, vehicles travel much longer in the range of $0.1<M/(N/logN)<0.6$ in the polymer-BPR configurations, even longer than that in the configurations identified by the polymer routing algorithm; such longer paths lead to the lower $\cost$ found by the polymer-BPR algorithm as we have seen in Figure~\ref{fig:sim}(a). These results suggest that vehicles may benefit most from the saved traveling time in the intermediate range values of $M$, similar to the findings in \cite{yeung2013physics}.

\begin{figure*}
\centering\includegraphics[height=3in]{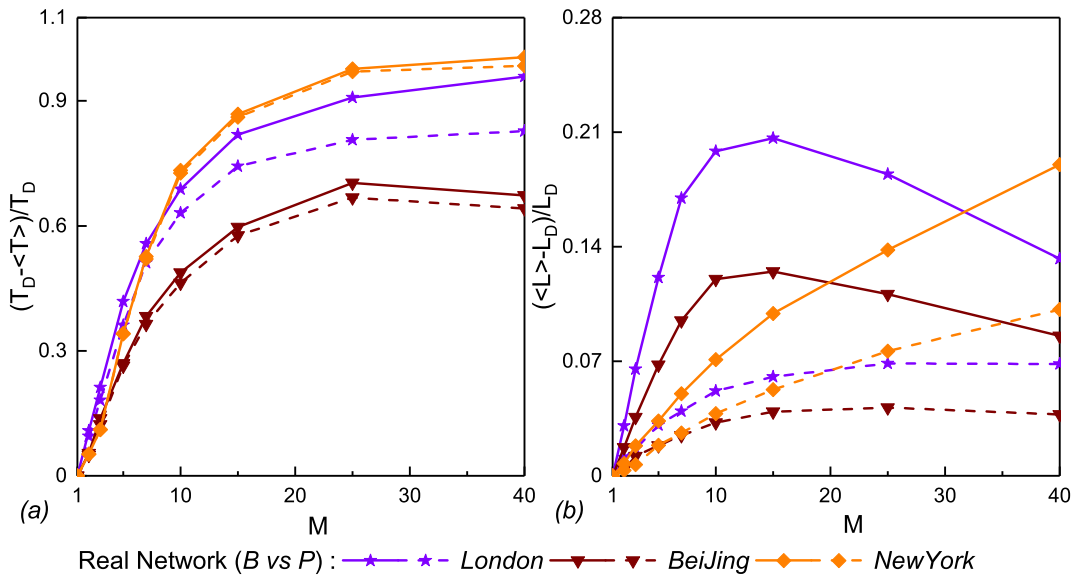}
\caption{
(a) The fractional difference $(\langle L\rangle-L_{D})/L_{D}$ in the path length, and (b) the fractional difference $(\cost_{D}-\langle \cost\rangle)/\cost_{D}$ in the cost $\cost$, between the path configurations identified by the polymer-BPR algorithm (as well as the polymer routing algorithm with $\gamma=2$) and the Dijsktra algorithm, as a function of $M/(N/\log N)$ which corresponds to the re-scaled number of vehicles. All simulation results are averaged over 2000 realizations of the London Underground network, the Beijing Rail Transit network and the New York road network.
}
\label{fig:real}
\end{figure*}

We also remark that the messages in the polymer-BPR do not always converge as $M$ increases, and various tricks suggested in \cite{saad2014physics} can be employed to improve the algorithmic convergence. Nevertheless, these tricks may also result in suboptimal solution.

To further demonstrate the effectiveness of our polymer-BPR algorithm in saving traveling time on real networks including the London Underground network, the New York Road network and the Beijing Rail Transit network, we show in Table~\ref{tab:energyDiff} the fractional difference in the estimated traveling time and the path length between the traffic configuration optimized by the various algorithms. Similar to the results on random regular graphs, vehicles travel via the polymer-BPR configuration saved an average of $70\%-80\%$ of traveling time compared to that of the shortest path configurations from the Dijsktra algorithm, in the expense of longer traveling distance by only $10\%-20\%$. Compared with the polymer routing algorithm with $\gamma=2$, the polymer-BPR also saves roughly $17\%$ of the traveling time on the London and the Beijing network in the expense of increased traveling distance. Examples of the identified traffic flow pattern by the polymer routing algorithm and the polymer-BPR algorithm are shown in \fig{fig:ld}(a) and (b) respectively. Nevertheless, such advantage is not observed on the New York Road network, probably because of the difficulty in algorithmic convergence.

The fractional difference between the estimated traveling time and the path length obtained by the polymer-BPR and the polymer routing algorithms are compared to those by the Dijsktra algorithm in Figure~\ref{fig:real}, as a function of the number of vehicles $M$. Similar to the results on random regular graph, both the polymer-BPR and the polymer routing algorithm outperform the Dijsktra path configurations in reducing the total travel time estimated by the BPR function, in the expense of an average traveling distance longer than that of the shortest path. The polymer-BPR algorithm also outperforms the the polymer routing algorithm in finding a shorter traveling time on the London and the Beijing networks for all the studied values of $M$, except on the New York Road network where the polymer-BPR and polymer routing algorithms have similar performance.

\section{Conclusion}

In this paper, we compared the cases where vehicles travel independently through their respective shortest path to the cases where vehicle paths are coordinated to suppress overloaded roads, and showed that the global vehicle traveling time as estimated by the link congestion function developed by the Bureau of Public Roads (BPR) can be greatly reduced if the paths of vehicles are coordinated. We incorporated the BPR function into the polymer routing algorithm developed in \cite{yeung2013physics} to identify optimal path configurations which minimize the total traveling time of vehicles. Through optimally coordinating vehicle path, our simulation results on generated and real transportation networks showed that our polymer-BPR algorithm outperforms the Dijsktra and the original polymer routing algorithm in finding a shorter global traveling time.



\bibliographystyle{named}
\bibliography{references}

\begin{thebibliography}{}

\bibitem[\protect\citeauthoryear{Chardaire \bgroup \em et al.\egroup
  }{2005}]{chardaire2005solving}
Pierre Chardaire, Geoff~P. McKeown, S.~A. Verity-Harrison, and S.~B.
  Richardson.
\newblock Solving a time-space network formulation for the convoy movement
  problem.
\newblock {\em Operations Research}, 53(2):219--230, 2005.

\bibitem[\protect\citeauthoryear{Dijkstra}{1959}]{dijkstra1959note}
Edsger~W Dijkstra.
\newblock A note on two problems in connexion with graphs.
\newblock {\em Numerische Mathematik}, 1(1):269--271, 1959.

\bibitem[\protect\citeauthoryear{Galina and Sys{\l}o}{1988}]{galina1988some}
Henryk Galina and Maciej~M. Sys{\l}o.
\newblock Some applications of graph theory to the study of polymer
  configuration.
\newblock {\em Discrete Applied Mathematics}, 19(1-3):167--176, 1988.

\bibitem[\protect\citeauthoryear{Grunitzki and
  Bazzan}{2016}]{grunitzki2016combining}
Ricardo Grunitzki and Ana~L.C. Bazzan.
\newblock Combining car-to-infrastructure communication and multi-agent
  reinforcement learning in route choice.
\newblock In {\em IJCAI'16}, 2016.

\bibitem[\protect\citeauthoryear{Liu \bgroup \em et al.\egroup
  }{2011}]{liu2011large}
Tian Liu, Xiaxiang Lin, Chaoyi Wang, Kaile Su, and Ke~Xu.
\newblock Large hinge width on sparse random hypergraphs.
\newblock In {\em IJCAI'11}, volume 2011, pages 611--616, 2011.

\bibitem[\protect\citeauthoryear{M{\'e}zard \bgroup \em et al.\egroup
  }{1987}]{mezard1987spin}
Marc M{\'e}zard, Giorgio Parisi, and Miguel Virasoro.
\newblock {\em Spin glass theory and beyond: An Introduction to the Replica
  Method and Its Applications}, volume~9.
\newblock World Scientific Publishing Company, 1987.

\bibitem[\protect\citeauthoryear{of Public~Roads}{1964}]{bpr1964traffic}
Bureau of~Public~Roads.
\newblock {\em Traffic Assignment Manual for Application with a Large, High
  Speed Computer}.
\newblock University of Michigan Library, 1964.

\bibitem[\protect\citeauthoryear{Saad \bgroup \em et al.\egroup
  }{2014}]{saad2014physics}
David Saad, Chi~H. Yeung, Georgios Rodolakis, Dimitris Syrivelis, Iordanis
  Koutsopoulos, Leandros Tassiulas, Rudiger Urbanke, Paolo Giaccone, and Emilio
  Leonardi.
\newblock Physics-inspired methods for networking and communications.
\newblock {\em IEEE Communications Magazine}, 52(11):144--151, 2014.

\bibitem[\protect\citeauthoryear{Sanghavi \bgroup \em et al.\egroup
  }{2009}]{sanghavi2009message}
Sujay Sanghavi, Devavrat Shah, and Alan~S. Willsky.
\newblock Message passing for maximum weight independent set.
\newblock {\em IEEE Transactions on Information Theory}, 55(11):4822--4834,
  2009.

\bibitem[\protect\citeauthoryear{Song \bgroup \em et al.\egroup
  }{2016}]{song2016deeptransport}
Xuan Song, Hiroshi Kanasugi, and Ryosuke Shibasaki.
\newblock Deeptransport: Prediction and simulation of human mobility and
  transportation mode at a citywide level.
\newblock In {\em IJCAI'16}, volume~16, pages 2618--2624, 2016.

\bibitem[\protect\citeauthoryear{Stefanello \bgroup \em et al.\egroup
  }{2016}]{stefanello2016using}
Fernando Stefanello, Bruno~C. da~Silva, and Ana~L.C. Bazzan.
\newblock Using topological statistics to bias and accelerate route choice:
  Preliminary findings in synthetic and real-world road networks.
\newblock In {\em IJCAI'16}, 2016.

\bibitem[\protect\citeauthoryear{Swoboda and Andres}{2017}]{swoboda2017message}
Paul Swoboda and Bjoern Andres.
\newblock A message passing algorithm for the minimum cost multicut problem.
\newblock In {\em CVPR'17}, volume~3, 2017.

\bibitem[\protect\citeauthoryear{Wu \bgroup \em et al.\egroup
  }{2006}]{wu2006transport}
Zhenhua Wu, Lidia~A. Braunstein, Shlomo Havlin, and H.~Eugene Stanley.
\newblock Transport in weighted networks: partition into superhighways and
  roads.
\newblock {\em Physical Review Letters}, 96(14):148702, 2006.

\bibitem[\protect\citeauthoryear{Xie \bgroup \em et al.\egroup
  }{2004}]{xie2004self}
Haiyong Xie, Lili Qiu, Yang~R. Yang, and Yin Zhang.
\newblock On self adaptive routing in dynamic environments-an evaluation and
  design using a simple, probabilistic scheme.
\newblock In {\em ICNP'04}, pages 12--23. IEEE, 2004.

\bibitem[\protect\citeauthoryear{Xu \bgroup \em et al.\egroup
  }{2018}]{xu2018effective}
Jie Xu, Yong Zhang, and Chunxiao Xing.
\newblock An effective selection method for vehicle alternative route under
  traffic congestion.
\newblock In {\em ICCT'18}, pages 494--499. IEEE, 2018.

\bibitem[\protect\citeauthoryear{Yang and
  Nikolova}{2016}]{yang2016approximation}
Ger Yang and Evdokia Nikolova.
\newblock Approximation algorithms for route planning with nonlinear
  objectives.
\newblock In {\em IJCAI'16}, pages 3209--3217, 2016.

\bibitem[\protect\citeauthoryear{Yeung and Saad}{2012}]{yeung2012competition}
Chi~H. Yeung and David Saad.
\newblock Competition for shortest paths on sparse graphs.
\newblock {\em Physical Review Letters}, 108(20):208701, 2012.

\bibitem[\protect\citeauthoryear{Yeung and Saad}{2013}]{yeung2013networkinga}
Chi~H. Yeung and David Saad.
\newblock Networking-a statistical physics perspective.
\newblock {\em Journal of Physics A: Mathematical and Theoretical},
  46(10):103001, 2013.

\bibitem[\protect\citeauthoryear{Yeung \bgroup \em et al.\egroup
  }{2013}]{yeung2013physics}
Chi~H. Yeung, David Saad, and K.~Y.~Michael Wong.
\newblock From the physics of interacting polymers to optimizing routes on the
  london underground.
\newblock {\em Proceedings of the National Academy of Sciences},
  110(34):13717--13722, 2013.

\bibitem[\protect\citeauthoryear{Zhang \bgroup \em et al.\egroup
  }{2017}]{zhang2017optimal}
Youzhi Zhang, Bo~An, Long Tran-Thanh, Zhen Wang, Jiarui Gan, and Nicholas~R.
  Jennings.
\newblock Optimal escape interdiction on transportation networks.
\newblock In {\em IJCAI'17}, 2017.

\bibitem[\protect\citeauthoryear{Zhu \bgroup \em et al.\egroup
  }{2006}]{zhu2006dynamic}
Yong Zhu, Constantinos Dovrolis, and Mostafa Ammar.
\newblock Dynamic overlay routing based on available bandwidth estimation: A
  simulation study.
\newblock {\em Computer Networks}, 50(6):742--762, 2006.

\end{thebibliography}

\end{document}